\documentclass[pra,twocolumn,showpacs,footinbib]{revtex4}
\usepackage{graphicx,rotating}

\newcommand{\be}{\begin{equation}}
\newcommand{\ee}{\end{equation}}
\newcommand{\bea}{\begin{eqnarray}}
\newcommand{\eea}{\end{eqnarray}}
\newcommand{\ba}{\begin{array}}
\newcommand{\ea}{\end{array}}

\begin{document}
\title{Excited states dynamics in time-dependent density functional theory:
      high-field molecular dissociation and harmonic generation.}

\author{Alberto Castro$^{1,2,3}$, M. A. L. Marques$^2$, Julio A. Alonso$^1$, George F. Bertsch$^3$ 
        and Angel Rubio$^2$}
\affiliation{
$^1$Departamento de F\'{\i}sica Te\'{o}rica, Universidad de Valladolid, 47011 Valladolid (Spain)\\
$^2$Departamento de F\'{\i}sica de Materiales, Facultad de Qu\'{\i}micas, Universidad del Pa\'{\i}s Vasco;\\
    Centro Mixto CSIC-UPV/EHU and Donostia International Physics Centre (DIPC), 20080 San Sebasti\'{a}n (Spain)\\
$^3$Physics Department and Institute for Nuclear Theory, University of Washington, Seattle WA 98195 (USA)}


\def\be{\begin{equation}}
\def\ee{\end{equation}}

\begin{abstract}
We present a theoretical description of femtosecond laser induced dynamics
of the hydrogen molecule and of singly ionised sodium dimers, 
based on a real-space, real-time,
implementation of time-dependent density functional theory (TDDFT). 
High harmonic generation, Coulomb explosion and 
laser induced photo-dissociation are observed. 
The scheme also describes non-adiabatic effects, such as the appearance of even harmonics
for homopolar but isotopically asymmetric dimers,
even if the ions were treated classically.
This TDDFT-based method is reliable, scalable, and extensible
to other phenomena such as photoisomerization,
molecular transport and chemical reactivity.
\end{abstract}

\maketitle

It is now possible to study electron and molecular dynamics in 
real time 
using various experimental techniques employing
intense ultra-short laser sources \cite{pBrabec00_yotros}.
Some examples of such investigations
include X-ray photoelectron spectroscopy of molecules \cite{pLeone00_yotros},
pump-probe ionisation measurements \cite{faraday_discussions}, production of high harmonics
as a source of soft X-rays \cite{x-ray-citations}, the measurement of 
electron-phonon interactions
in thin films \cite{pProbst97}, and the estimation of the onset of Coulomb screening 
\cite{pHuber01}.
A technologically important and very active field of research is the application of ultra-short laser
pulses to induce, control and monitor
chemical reactions \cite{pDion96_yotros,pGerber01,pYamanouchi02}. 
Whenever the intensity of the laser field is comparable
to the molecular electronic fields, perturbative 
expansions break down and new processes appear, 
which are not fully understood from a microscopical point of view \cite{new_processes}.
A practical and accurate computational framework to descibe excited-state electron-ion dynamics
is therefore still needed.

Not surprisingly, the smallest systems have attracted particular attention 
from both experimentalists and theoreticians, as a bench-horse to improve
our understanding of electron dynamics at the femtosecond scale \cite{pKnight97, atoms_and_dimers_studies}.
However, the methods used in these calculations
can not be easily extended to larger and more realistic systems. The exact quantum
mechanical solution of a 3D system of more than three particles
is certainly not feasible with state-of-the-art computers. 
1D models are much easier to handle, but they can not really be used as predictive
tools for problems involving the interaction of lasers with large clusters
or solid-state systems of technological relevance.

To tackle such a problem, time-dependent density functional 
theory 
(TDDFT) \cite{tddft} appears as a valuable tool.
Even with the simplest approximation to the exchange-correlation potential, the
adiabatic local density approximation (ALDA), one obtains a very good compromise
between computational ease and accuracy \cite{ALDA_problems}.
TDDFT can certainly be applied to large systems
in non perturbative regimes, while providing a consistent treatment
of electron correlation. It has been well tested in the study
of electron excitations, like the optical absorption spectra in
the linear regime \cite{pYabana96, review}.
Although almost all applications of TDDFT in the field of laser physics
have only involved electronic dynamics, recent attempts 
have also been made at describing the coupled nuclear and electronic 
motion in laser fields \cite{pCalvayrac00_yotros}, accounting for the nuclear motion classically.
A full quantum mechanical treatment of the system
could in principle be done within
a multi-component TDDFT,
although it has not been tried for
more than three particles \cite{pKreibich01-2}.
However, since many vibrational quanta are coherently excited, there is good motivation
for the classical treatment of the nuclei.
The purpose of this work is to illustrate a general method
to study many-electron systems subject to strong laser fields.
It is based on the quantum mechanical propagation of the electronic
wave packet -- described within TDDFT --
combined with classical motion of the nuclei.
As the laser field populates the excited Born-Oppenheimer
surfaces, this scheme includes diabatic effects, while maintaining
a good scaling with the size of the system.
As an illustration
we focused on one and two electron dimers, namely
Na$_2^+$ and the hydrogen molecule.

The equations of motion may be derived from the Lagrangian:
\def\a{\alpha}
\begin{eqnarray}
  \nonumber
  {\cal L} = 
  \sum_\a \left[ {1\over 2} m_\a  \dot{\vec{R}}_\a^2
  + Z_\a \;\vec R_\a \cdot \vec {\cal E} (t) \right]
  - \sum_{\a<{\a'}} {Z_\a Z_{\a'} \over | \vec{R}_\a - \vec{R}_{\a'}|}
\\
  \sum_i \langle \phi_i |
  i\;\hbar{\partial \over \partial t}
  -e \; \vec x \cdot \vec {\cal E} (t)
  |\phi_i\rangle
  - E_{DFT} (\{\phi\}, \{\vec R\})
  \,,
\end{eqnarray}
where $E_{DFT}$ is the usual Kohn-Sham density functional, depending on the
electron orbitals $\{\phi\}$ and the nuclear coordinates $\{\vec R\}$, and $\vec{\cal E}(t)$
is the time-dependent electric field from the laser pulse. Variation of the
Lagrangian then yields Newton's equations for the nuclear coordinates,
\begin{eqnarray}
  \nonumber
  m_\a {d^2 \vec{R}_\a \over d t^2} & = & - \vec{\nabla}_{R_\a} E_{DFT}
  \\ \label{eq:forces}
  & & -\sum_{\a'}{(\vec {R}_\a
  -\vec R_{\a'})Z_\a Z_{\a'}\over |\vec{R}_\a -\vec{R}_{\a'}|^3} + Z_\a \; \vec {\cal E} (t),  
\end{eqnarray}
and the usual TDDFT equations for the orbital variables \cite{tddft}.
We solved these equations in real time, following the method of
Yabana and Bertsch \cite{pYabana96},  using a real-space grid
representation of the orbitals \cite{pFlocard78,pChelikowsky94_yotros}. This scheme has the advantage
that the Kohn-Sham Hamiltonian is a very sparse matrix.  The forces in Eq. (\ref{eq:forces})
are calculated with the help of a generalised Hellmann-Feynman theorem,
\be
 - \vec{\nabla}_{\vec{R}_\a} E_{DFT}= -\sum_i \langle \phi_i |\vec{\nabla}_\a H_{KS}|\phi_i\rangle.
\ee
where $H_{KS} = \delta E_{DFT} / \delta n({\bf r},t)$ is the Kohn-Sham
Hamiltonian.
For numerical reasons we represent the electron-ion interaction by
norm-conserving non-local Troullier-Martins pseudopotentials 
\cite{pTroullierMartins91}. 

\begin{figure}
\includegraphics[scale=0.33]{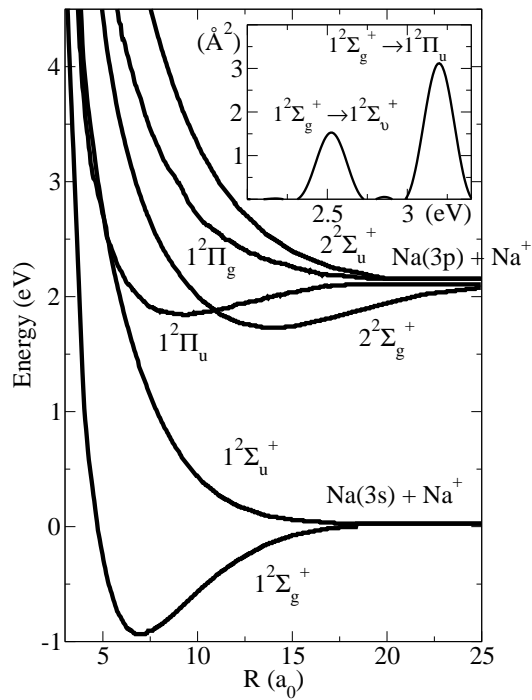}
\caption{\label{fig:pes}
Adiabatic energy surfaces of the Na$_2^{+}$ dimer,
as obtained by our three-dimensional real-space code.
Similar results can be found in Ref. \cite{pMagnier_Oise96}.
In the inset, the computed photoabsorption cross-section of the
same molecule.}
\end{figure}

We have studied two different classes of time-dependent problems:
photofragmentation and high harmonic generation.
We now discuss the first case, the Na$_2^+$ dimer in 
a femtosecond laser field.  
It is a good test, since it has been exhaustively studied using
a diverseness of approaches \cite{pMagnier99,pMagnier96_yotras,pGerber01}.
In particular, a recent experiment \cite{pGerber01} focused on
the photofragmentation of Na$_2^+$ in intense femtosecond laser fields.
Using a pump-probe technique, the authors discovered that Na$_2^+$
dissociated in four different channels, ranging from simple field ionisation
followed by Coulomb explosion, to photodissociation on light-induced 
potentials.
For these calculations, we used an uniform grid
spacing of 0.3~\AA, and the system was confined to a sphere of
radius 10~\AA. 
In these one-electron calculations, we omitted the core-valence exchange-correlation.
In this case the total electronic energy is given by the Kohn-Sham eigenvalues,
and they can be used to compute the adiabatic potential energy surfaces
shown in Fig. \ref{fig:pes}. 
The two lowest single-photon transitions from the $1^2\Sigma_g^+$ ground state
are at 2.5\,eV (to the $1^2\Sigma_u^+$ state) and at 3.2\,eV (to the $1^2\Pi_u$ state).
The latter is achieved by a laser polarized perpendicularly to the internuclear axis.
These energies accord well with the observed
single-photon transitions \cite{bonacic}. For the time-dependent calculations,
we start with the dimer in its ground state, which is then propagated with
a modified Krank-Nicholson scheme \cite{pFlocard78}.
The time step for the time integration was 
0.005 $\hbar$ eV$^{-1} \approx 0.003$ fs.  A simple check on
the implementation of the time evolution operator consists of calculating
the linear photo-absorption spectrum, using a weak 
$\delta$-function external field, as in Ref. \cite{pYabana96}.  Almost all the
spectral weight is concentrated in two peaks (see inset in Fig.~\ref{fig:pes}),
which are at energies
corresponding exactly to the vertical transitions between energy
surfaces.  We note that this exact correspondance is only obtained
for one-electron systems: in general the TDDFT spectra will have
shifts from the energy surfaces determined by the Kohn-Sham eigenvalues.
\begin{figure*}[t]
\includegraphics[scale=0.43]{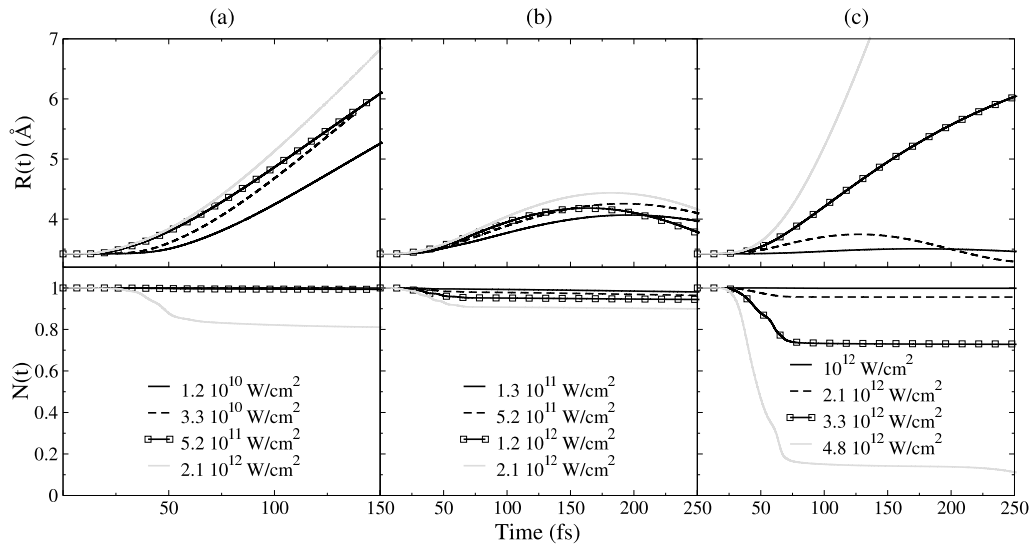}
\caption{\label{fig:resonant_dissociation}
Evolution of internuclear distance (top panel) and
electronic charge in simulation region (bottom panel) for
the Na$_2^+$ molecule. The dimers are excited with
laser pulses of 2.5, 3.2 and
1.57~eV in columns (a), (b) and (c) respectively.
}
\end{figure*}
\begin{figure}[b]
\includegraphics[scale=0.32]{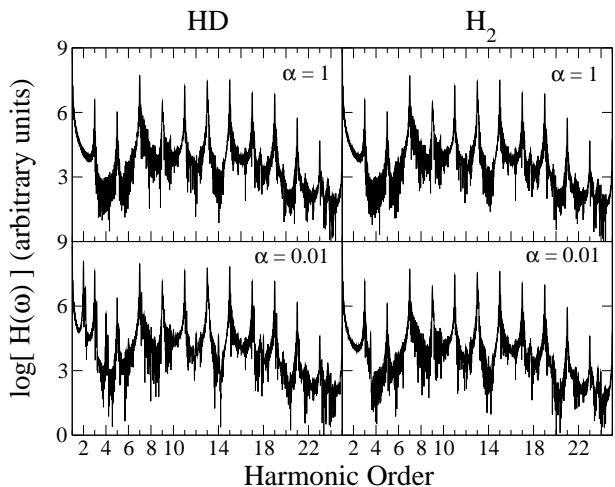}
\caption{\label{fig:harmonics.ps}
Harmonic spectra of HD (left panels) and H$_2$ (right panels). 
The nuclear masses used in the calculation are $m'=\alpha m$, being
$m$ the real mass. In this way, top
plots were made using for the nuclear masses their real values
whereas bottom plots were made using a hundredth of their real values.
}
\end{figure}

Next we examine the evolution of the dimer under high-field excitation.
We consider external fields of the form:
\begin{equation}
 \vec {\cal E}(t) = \left( \frac{8\pi}{c} I_0\right)^{(1/2)} 
 \sin \left( \pi \; \frac{t}{\tau} \right) \sin(\omega t) {\hat e},
 \;\;\;
 0 < t <\tau,  
\end{equation}
where $I_0$ is the maximum intensity of the pulse,
${\hat e}$ is the polarization vector,
and $\tau$ is the 
pulse duration, taken as $\tau = 80 $ fs.  As a first case, we
examine the effect of excitation at the lower resonant frequency,
$\omega = 2.5$ eV.  In Fig. 2(a), we present a series of
runs at different intensities, ranging from weak ($10^{10}$ W/cm$^2$)
to moderate ($2.1\times 10^{12}$ W/cm$^2$).  Since the $1^2\Sigma_u^+$
surface is anti-bonding, excitation at this resonant frequency should 
lead to dissociation, even at moderate intensities.  This is indeed confirmed
by our calculations.
The upper panel depicts the internuclear separation 
of the dimer, which exhibits an acceleration during the laser pulse and
a nearly constant velocity expansion thereafter.  Clearly, the dimer
dissociates at all field levels that we applied.  To examine the ionisation
of the dimer,
we assumed that any density reaching the edges of the simulation box corresponds
to unbound electrons. By absorbing this density at the boundaries, we can thus
define the ionisation probability as $I(t) = 1 - N(t)$, where $N(t)$ is the charge
that remains inside the simulation box at time $t$. The lower panel of Fig. 2(a) shows $N$ as a 
function of $t$.
We see that there is practically no ionisation
for the lower fields, and only a 20\% ionization probability for the 
$2.1 \times 10^{12}$ W/cm$^2$ field.  Thus, in this range of intensities, the
laser dissociates the dimer without ionising it.

We next consider the excitation at the upper resonance frequency, 
$\omega = $ 3.2\,eV, corresponding
to an electric field perpendicular to the dimer axis. 
Since the $1^2\Pi_{u}$ surface is bonding, 
[see Fig.~\ref{fig:resonant_dissociation}~(b)]
no dissociation
is expected unless the Coulomb explosion channel is opened through
ionisation. We see that the dimer remains bound over the entire range
of intensities that produced dissociation at the lower resonant frequency.

Finally, we also performed simulations at the
non-resonant frequency $\omega =$ 1.57\,eV,
the one used in Ref.
\cite{pGerber01}. Fig.~\ref{fig:resonant_dissociation}~(c)
shows how dissociation now occurs only at much higher intensities,
and it is mainly due to ionisation (almost absent in the resonant
calculations for the range of intensities used): 
since Na$_2^{2+}$ has no bonding states, ionisation is followed by 
Coulomb explosion.

Another process in which the nuclear motion may play an important role is high harmonic
generation.  Even harmonics may be created by irradiating HD with an intense laser pulse,
but not by irradiating H$_2$:  even harmonic generation
is forbidden for a centrosymmetric molecule.
In an adiabatic
treatment of the nuclear coordinates, the nuclear masses play no role and
the even harmonics can not appear.
This is no longer the case if non-adiabatic effects are taken into account, for the 
different masses of H and D break the symmetry.
Kreibich et al.  \cite{pKreibich01} studied this  process in 
a 1-D model with a full quantum mechanical treatment of the nuclear
motion, finding strong even harmonics at high
harmonic number.  To discern whether the classical treatment of nuclear
motion also produces these harmonics, 
we studied the same 1D problem within our framework.
As in Ref.~\cite{pKreibich01},
we took the laser field to have a frequency 
of 1.6 eV, and an intensity that rises linearly to $10^{14}$ W/cm$^2$
over an interval of 10 optical cycles, and is held constant thereafter.
We then calculated the spectral intensity of the generated harmonics, $H(\omega)$:
\be \label{eq:harmonic_spectrum}
  H(\omega) \sim \left|\int\!\!d t\: e^{i\omega\: t}\:
  \frac{d^2}{d t^2} \left<\Psi(t)\right|\hat{e} \cdot \vec D \left|\Psi(t)\right>
  \right|^2\,.
\ee
We find that the
classical treatment does indeed produce even harmonics, but
much smaller than the quantum treatment.  The results
are shown in Fig. \ref{fig:harmonics.ps}.  The top left panel depicts
the harmonic spectrum for HD, and only odd harmonics are apparent.
However, it may be proved that the HD Hamiltonian already violates centrosymmetry
within our classical treatment, through a term of the form:
$$-\frac{1}{2}\left(\frac{1}{M_H}-\frac{1}{M_D}\right)P(t)\left(\hat{p}_1 + 
\hat{p}_2\right),
$$
where $P(t)=\frac{1}{2}\left(P_H(t)-P_D(t)\right)$ is the relative time-dependent
nuclear momentum
and $\hat{p}_i$ are the 
electronic momentum operators \cite{ref_frame}.  Its effect can
be enhanced by decreasing the nuclear masses.  In the bottom left
panel, the H and D masses have been decreased by a factor 100, and
then the second- and fourth-order harmonics become visible.  As a
qualitative check of the numerics, we also show the same graphs for
H$_2$, in which no even harmonics can occur.

Thus we see that on qualitative level the non-adiabatic dynamics generating
even harmonics are obtained with the classical treatment of the
nuclear coordinates. However, the quantum treatment may
be needed for a quantitative result.  
By describing the nuclei quantum mechanically,
the ground 
state violates centrosymmetry and the even harmonics can be generated
even if the nuclear motion is frozen.  In contrast, in the classical
treatment the ground state is symmetric and the symmetry violation only builds
up as the nuclei move.

In summary, we have examined the computational feasibility of
including nuclear dynamics in time-dependent density functional
theory 
using
a pseudopotential code to study the femtosecond laser
induced dynamics of sodium dimers. 
Using this approach for treating the Na$_2^+$ dimer, 
we were able to distinguish different phodissociation regimes, ranging
from dissociation on light induced potentials, to field ionisation 
followed by Coulomb explosion. 
Electronic and ionic degrees of freedom are thus coupled, so that one can
observe the electron-phonon transfer of energy.
We also found, with another example, that non-adiabatic effects are 
present in the general treatment based on Eq. (1).
One of the major attractivenesses 
of this method resides in its reasonable
scaling behaviour when applied to larger systems.
We thus expect to be able to tackle problems like
photoisomerization or even photochemical reactivity in
systems of dozens of atoms in the near future.

This work was supported by the RTN program
of the European Union NANOPHASE (contract HPRN-CT-2000-00167),
Basque Country University,
Iberdrola S.A. and DGESIC (PB98-0345).
GB acknowledges support by the US Department of Energy under
Contract Nr. E-FG-06-90ER-411132.
Computer time was kindly provided by the CEPBA.
We thank E.~K.~U. Gross for enlightenment discussions.
AC thanks the University of Washington and the DIPC for kind hospitality.

\bibliographystyle{prsty}
\bibliography{abbrevs,biblio}

\begin{thebibliography}{10}
\bibitem{new_processes}
Examples commonly invoked are bond softening, vibrational population trapping,
molecular alignment and above threshold dissociation.
\bibitem{ref_frame}
A canonical transformation to the nuclear geometrical center has been performed.
\end{thebibliography}

\end{document}